\documentclass{moriond}

\def\be{\begin{equation}}
\def\ee{\end{equation}}
\def\bea{\begin{eqnarray}}
\def\eea{\end{eqnarray}}

\newcommand{\Photo}{}

\begin{document}
\vspace*{4cm}
\title{Cosmic Ray Physics with the KM3NeT Telescopes}

\author{ Br\'{i}an \'{O} Fearraigh, on behalf of the KM3NeT Collaboration}

\address{INFN – Sezione di Genova \\
Via Dodecaneso 33, Genova, 16146 Italy}

\maketitle\abstracts{
The KM3NeT research infrastructure instruments a large volume of seawater using photomultiplier
tubes, which are sensitive to the Cherenkov radiation stimulated by the products of neutrino
interactions in the water, as well as that stimulated by atmospheric muons which penetrate the
sea depths. The KM3NeT/ARCA and KM3NeT/ORCA detectors are situated at different depths in
the Mediterranean Sea, with different extension and densities of the photo-detection elements.
Although operating independently, taken as a whole the two detectors provide a wide energy
coverage for the atmospheric muons flux. Through the detection and analysis of these
atmospheric muons, a variety of physics studies are possible with the KM3NeT telescope. A
measurement of the atmospheric muon neutrino flux has been carried out with data from the
initial six detection units of the KM3NeT/ORCA detector. Relatedly to the atmospheric muon flux,
the recent atmospheric lepton model ‘Daemonflux’ has been incorporated into the KM3NeT
Monte Carlo event generator for atmospheric muon bundles. This has resulted in a stark
alleviation of the atmospheric muon data-Monte Carlo simulation discrepancy - a systemic issue
in cosmic ray experiments referred to as the ‘Muon Puzzle’ - and a comprehensive description of
the atmospheric muon data in KM3NeT. These atmospheric muons are also used in the calibration
of the detectors, as well as constraining systematic uncertainties in the detectors such as the
optical properties of the instrumented seawater. An overview of these topics, and other cosmic
ray analyses, is presented.}

\section{The KM3NeT experiment}
The Cubic Kilometre Neutrino Telescope, KM3NeT, is a deep-sea neutrino detection infrastructure operating from, and under construction at, two separate locations in the Mediterranean Sea~\cite{km3netloi}. At both sites, the seawater is being instrumented with an array of photo-sensors which are attached to a sea floor network of electro-optical cables. The photo-sensors are used to detect the Cherenkov radiation resulting from relativistic particles traversing the seawater. Specifically, 31 PMTs are contained within a glass sphere hosting electronics and environmental sensors, forming a Digital Optical Module (DOM)~\cite{dompaper}. Eighteen of these DOMs are arranged between two Dyneema ropes, alongside optical and power connections to a base module on the sea floor, forming a structure referred to as a Detection Unit (DU). The DU is held aloft through the buoyancy of the DOMs and additional buoys. 

These DUs are being deployed in phases, resulting in partial detector configurations. The inter-DU spacing on the sea floor, and the spacing between optical modules along the DU, is optimised to give two distinct detectors. The KM3NeT/ARCA detector, located $\sim$100 km offshore Portopalo di Capo Passero in Italy, and at a depth of $\sim$3.5 km below sea level, is a more sparse detector geometry optimised for the detection of TeV-PeV neutrinos which are used for neutrino astronomy studies. The KM3NeT/ORCA detector, located $\sim$40 km offshore Toulon in France and $\sim$2.5 km below sea level, is a more compact detector configuration for the detection of GeV-scale neutrinos used to measure the atmospheric neutrino oscillation parameters and determine the neutrino mass ordering. At the time of writing, 33 out of approximately 100 planned DUs have been deployed at the ORCA detector location, and 51 out of approximately 200 envisaged DUs at the ARCA detector site. Although the science goals of KM3NeT primarily are those involving neutrino detection, the atmospheric muons that are produced in extensive air showers reach the depths at which both detectors operate and can be studied for cosmic ray physics research.
\section{Cosmic ray physics \& KM3NeT}
Cosmic rays impinge on the Earth's atmosphere, resulting in extensive air showers containing muonic, hadronic, and electromagnetic components. The forward-boosted muons can reach the depths at which KM3NeT operates, and allow for the study of these air showers from which they came. With detectors at two different depths, KM3NeT is sensitive to a wide energy range of these muons. The primary cosmic ray energies to which KM3NeT is sensitive has been estimated using \texttt{CORSIKA}~\cite{corsika} simulations, showing TeV-PeV sensitivity~\cite{andreypaper}. %The same TeV-PeV energy range is also valid for the atmospheric muons at sea level, which are then detected by KM3NeT. %Please refer to Fig 2 for the primary cosmic ray energies in either detector for the different primary particle atoms.

\section{Overview of physics studies}
A summary of the cosmic ray and atmospheric lepton physics studies being carried out by the KM3NeT collaboration is presented below.

\subsection{Atmospheric muon neutrino flux measurement}
\begin{figure}[hb]
\centerline{\includegraphics[width=0.5\linewidth]{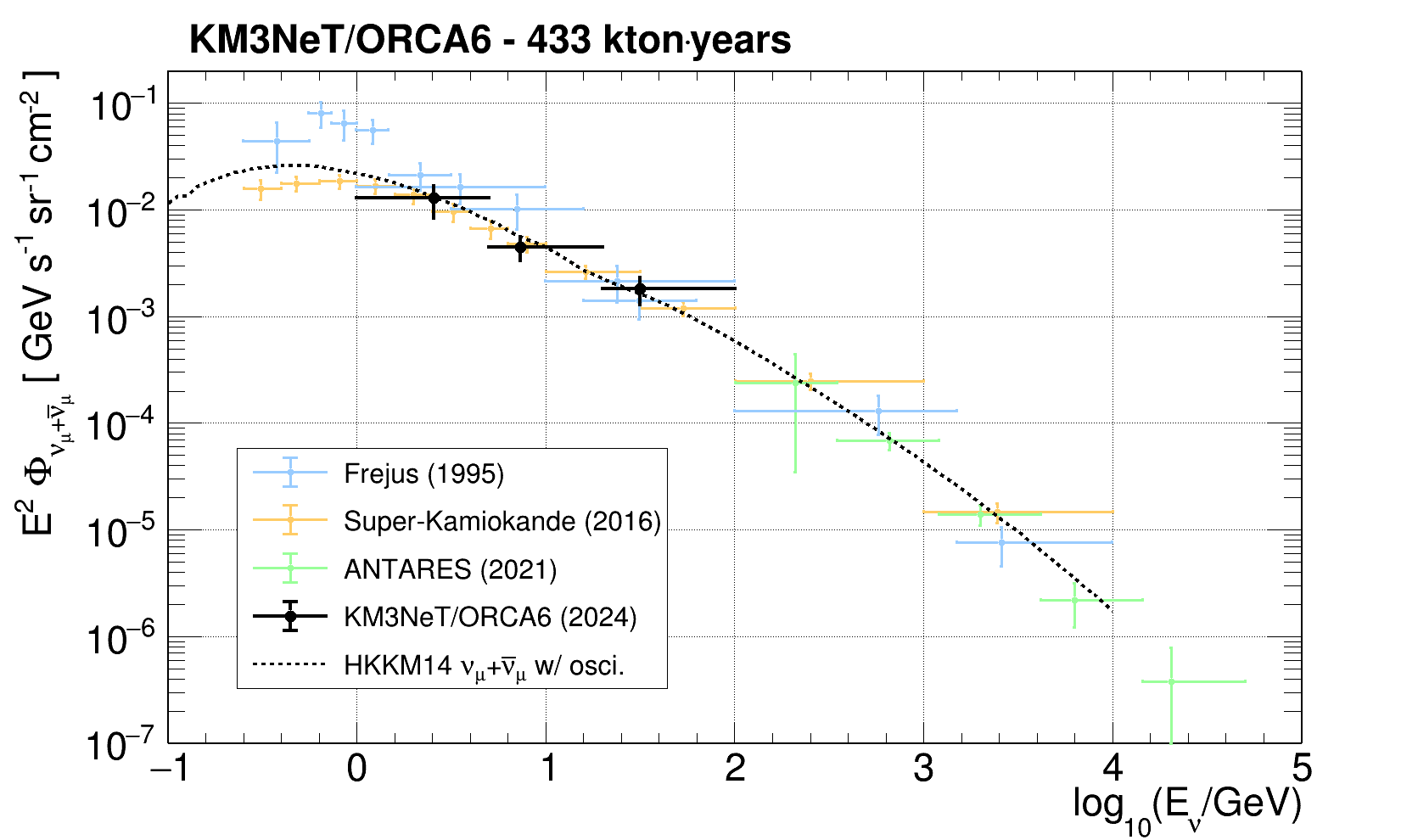}}
\caption[]{The atmospheric $\nu_{\mu}+\bar{\nu}_{\mu}$ flux measurement with the ORCA-6 detector, compared to measurements from other experiments.}
\label{fig:flux}
\end{figure}

A first measurement of the atmospheric $\nu_{\mu} + \bar{\nu}_{\mu}$ has been carried out with data from the ORCA-6 detector (i.e. with 6 operating detection units)~\cite{numuflux}. Using 433 kton-yr of data, an event selection is performed to select up-going neutrinos in the detector and separate them from the background. The energy spectrum is obtained from the data via an unfolding technique, and the flux then determined. The measured flux is shown in Fig.~\ref{fig:flux}, comparing the KM3NeT data to those from other experiments and model predictions. 

\subsection{Parameterisation of the atmospheric muon flux using Daemonflux}
\begin{figure}
\begin{minipage}{0.5\linewidth}
\centerline{\includegraphics[width=0.95\linewidth]{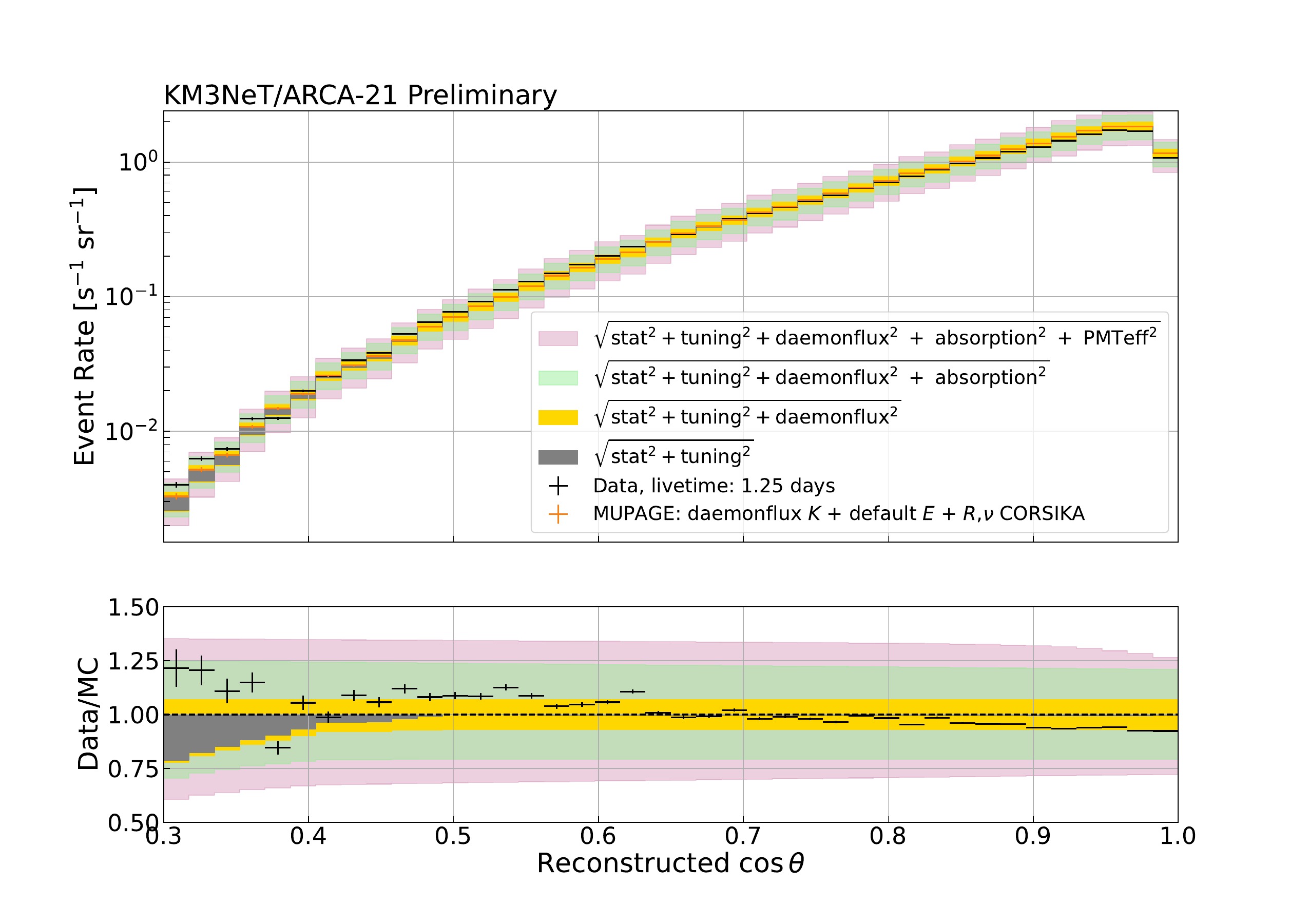}}
\end{minipage}
\begin{minipage}{0.5\linewidth}
\centerline{\includegraphics[width=0.95\linewidth]{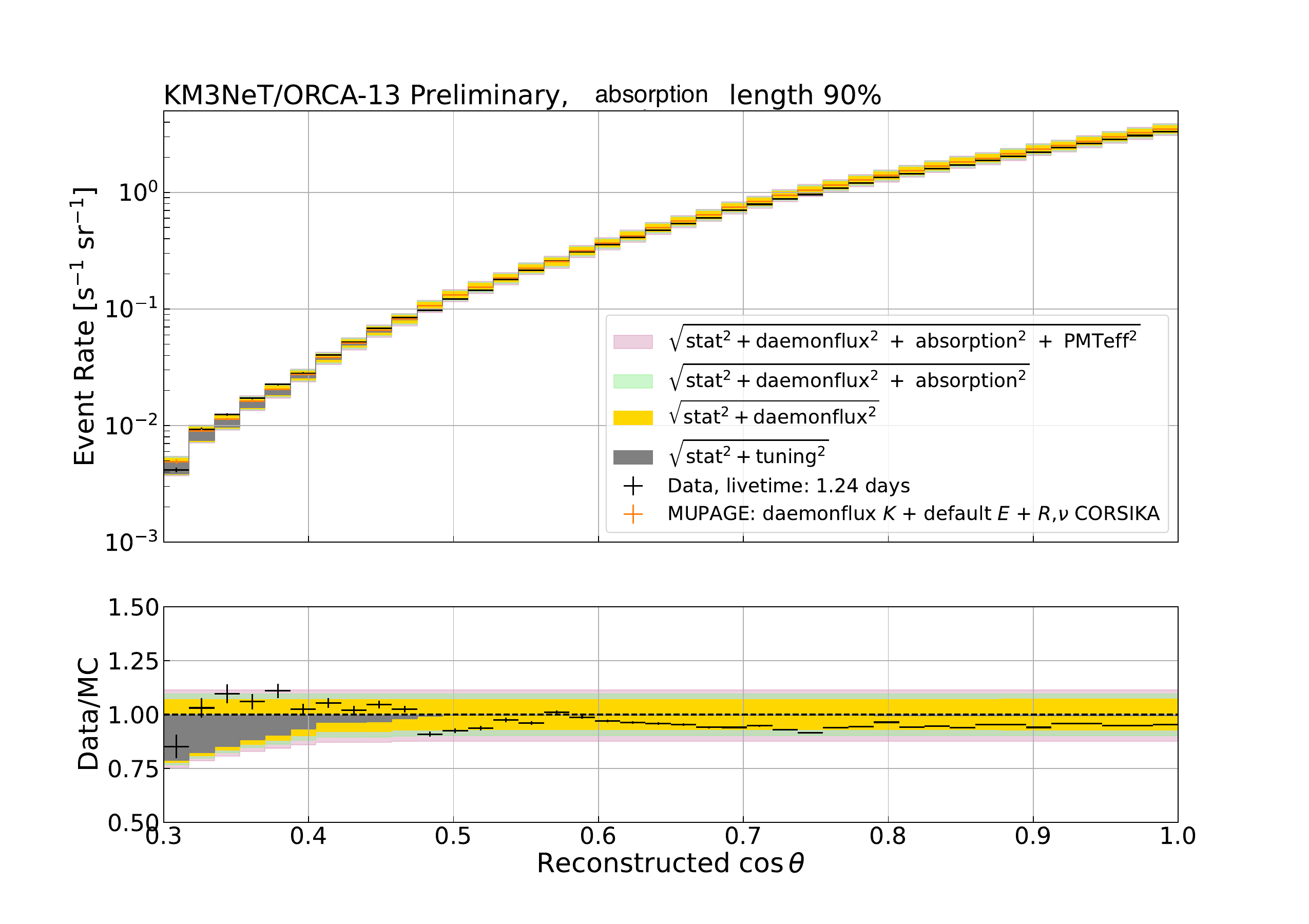}}
\end{minipage}
\caption[]{Data-simulation comparisons of the KM3NeT partial ARCA and ORCA detectors, comparing atmospheric muon data to the simulation for which the flux has been parameterised according to \texttt{Daemonflux}~\cite{tevpa}.}
\label{fig:daemonfluxdatamc}
\end{figure}

The \texttt{Daemonflux}~\cite{daemonflux} data-driven atmospheric lepton flux model has been incorporated into the muon bundle generator \texttt{MUPAGE}~\cite{mupage1,mupage2}, in which parametric formulae are used to describe the single and multi-muon bundle flux, energy distribution, and the lateral spread of muons. Using event weights which encapsulate the differences between \texttt{CORSIKA} and \texttt{Daemonflux}, the parametric equations of \texttt{MUPAGE} are fitted to the single muon zenith angle distributions from weighted \texttt{CORSIKA} simulations at different depths~\cite{icrc}. Combining this new parametric description of the single muon flux with a multiplicity parameterisation determined from (un-weighted) \texttt{CORSIKA} simulations~\cite{andreypaper}, has resulted in an alleviation in KM3NeT of the discrepancy between atmospheric muon data and the simulation -- a phenomenon reported by multiple atmospheric muon detectors, referred to as the Muon Puzzle. The KM3NeT data and updated muon flux parameterisation, incorporating the \texttt{Daemonflux} model, is shown in Fig.~\ref{fig:daemonfluxdatamc} for the ARCA-21 and ORCA-13 detectors~\cite{tevpa}. %An evaluation of the systematic uncertainties is included, with the flux uncertainty from \texttt{Daemonflux} accounting for the cosmic ray flux 

\subsection{Systematic uncertainty evaluation}
\label{optical}
Two of the primary systematic uncertainties affecting all KM3NeT studies are the light detection efficiency of the PMTs, and the knowledge of the absorption length of light in seawater. A technique using stopping muons has been developed to measure these quantities, whereby the well-understood light emission profile - with reduced stochasticity - of stopping muons in the ORCA-6 detector are taken advantage of~\cite{opticalproperties}. 
\begin{figure}[h]
\centerline{\includegraphics[width=0.4\linewidth]{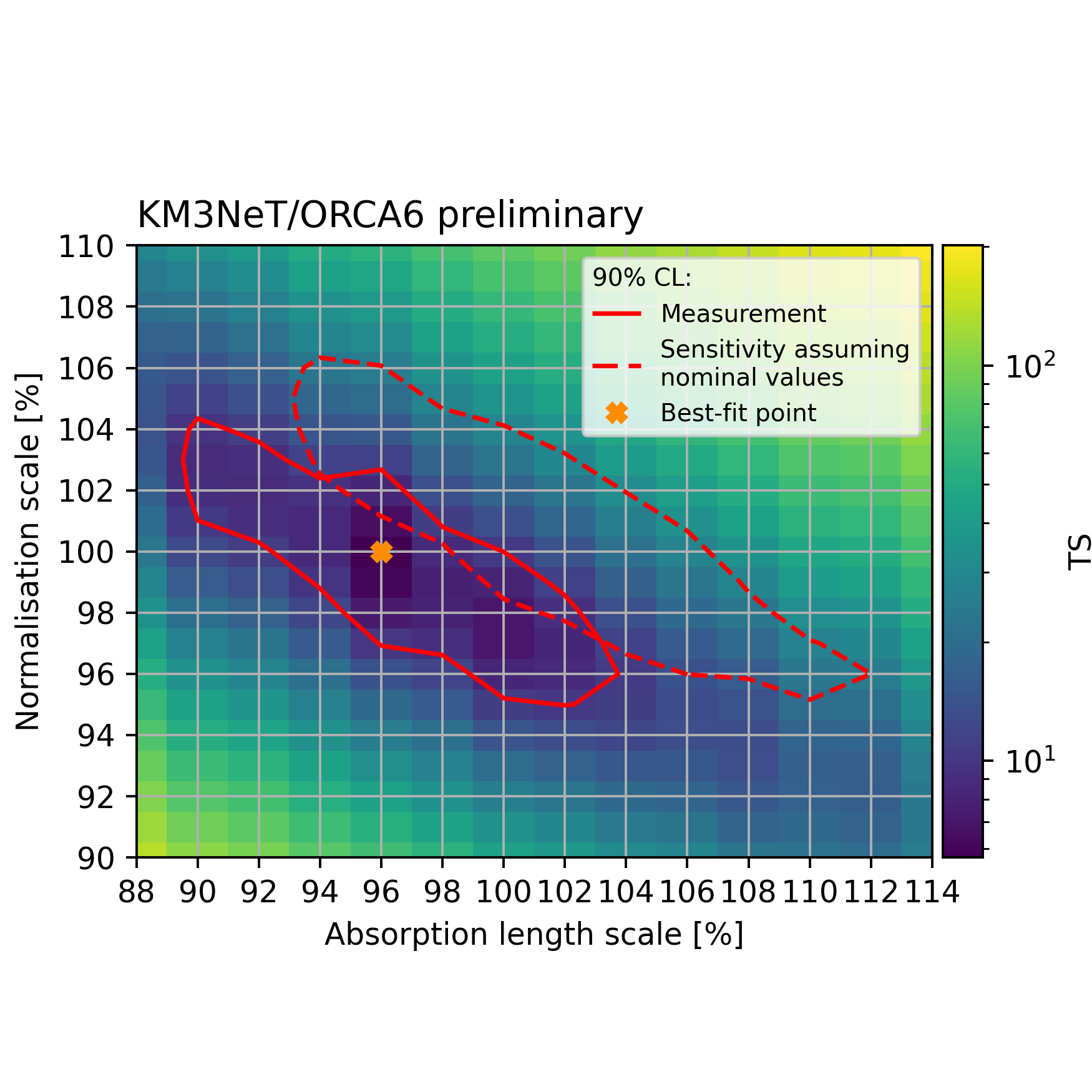}}
\caption[]{The determined values of the absorption length of light and the normalisation scale (i.e. light detection efficiency) using stopping muons in the ORCA-6 detector are depcited. The solid contour indicates the 90\% CL for the measurement. The best-fit point for the data is shown with the orange cross, and the dashed contour shows the expected sensitivity using the nominal values for the absorption length and normalisation~\cite{opticalproperties}.}
\label{fig:waterproperties}
\end{figure}
Probability density functions (PDFs) of light detection are estimated as a function of the distance of closest approach between the muon track and each PMT, given the reconstructed track direction and position. These PDFs are compared between data and simulations for the stopping muons, allowing for a measurement of the absorption length of light and the light detection efficiency (referred to as the normalisation). The results for the ORCA-6 detector are shown in Fig.~\ref{fig:waterproperties}.

\subsection{Cosmic ray anisotropy \& Prompt muon flux sensitivity}
A first study of the cosmic ray direction anisotropy has been made with data from the ARCA-21 detector. The East-West method is used to estimate the rate of reconstructed tracks from either direction, to disentangle the anisotropy from instrumental effects. The amplitude and phase of the first harmonic of the cosmic ray anisotropy is derived from directly fitting
the reconstructed track rate; a potential anisotropy signal is estimated with $\sim$2$\sigma$ and an upper limit derived on the first harmonic amplitude and phase is compared to other experiments~\cite{anisotropy}. 

Separately, an estimate of the sensitivity of KM3NeT to the prompt atmospheric muon flux has been made. This involves a selection of prompt muons in the atmospheric muon simulation, and a dedicated machine-learning based reconstruction of the muon bundle energy and multiplicity. From these observables, a sensitivity to the prompt flux is determined for the 6- and 115-DU detector configurations of ARCA and ORCA~\cite{prompt}. 

\subsection{Detector calibration}
Atmospheric muons are used in calibrating the KM3NeT detectors. Being a  well-defined signal, they may be used to estimate the relative timing offsets in the detector. They may also be used in estimating the pointing accuracy of the detectors. For example, the position of the Sun and the Moon shadows in the sky have been determined using ORCA-6 data to a high degree of accuracy: 6.2$\sigma$ and 4.2$\sigma$, respectively~\cite{sunmoonshadow}. These celestial bodies are expected to obstruct a number of cosmic rays reaching the Earth, thus resulting in a deficit of atmospheric muons measured from their (well-known) position in the sky. The demonstrated sensitivity to the two shadows reflects the good understanding of the detector positioning, orientation, time calibration, and reconstruction capabilities, using a limited data set with a partial detector configuration.

\section*{References}
\bibliography{moriond}

\end{document}